# Towards Determining Mechanical Properties of Brain-Skull Interface Under Tension and Compression

Sajjad Arzemanzadeh[1*][0000-0001-7381-1777], Benjamin Zwick[1][0000-0003-0184-1082], Karol Miller[1][0000-0002-6577-2082], Tim Rosenow[2][0000-0003-3727-5131], Stuart I. Hodgetts[3, 4][0000-0002-3318-0410], and Adam Wittek[1][0000-0001-9780-8361]

[1] Intelligent System for Medicine Laboratory (ISML), School of Mechanical Engineering, The University of Western Australia, Perth 6009, WA, Australia

[2] Western Australia National Imaging Facility, The University of Western Australia, Perth 6009, WA, Australia

[3] Spinal Cord Repair Laboratory, School of Human Sciences, The University of Western Australia, Perth 6009, WA, Australia

[4] Perron Institute for Neurological and Translational Science, Verdun Street, Nedlands, 6009, WA, Australia

**Abstract.** Computational biomechanics models of the brain have become an important tool for investigating the brain responses to mechanical loads. The geometry, loading conditions, and constitutive properties of such brain models are well-studied and generally accepted. However, there is a lack of experimental evidence to support models of the layers of tissues (meninges) connecting the brain with the skull (brain-skull interface), which determine boundary conditions for the brain. We present a new protocol for determining the biomechanical properties of the brain-skull interface and present the preliminary results (for a small number of tissue samples extracted from sheep cadaver heads) obtained using this method. The method consists of biomechanical experiments using brain tissue and brain-skull complex (consisting of the brain tissue, brain-skull interface, and skull bone) and comprehensive computer simulation of the experiments using the finite element (FE) method. Application of the FE simulations allowed us to abandon the traditionally used approaches that rely on analytical formulations that assume cuboidal (or cylindrical) sample geometry when determining the parameters that describe the biomechanical behaviour of the brain tissue and brain-skull interface. In the simulations, we used accurate 3D geometry of the samples obtained from magnetic resonance images (MRIs). Our results indicate that the behaviour of the brain-skull interface under compressive loading appreciably differs from that under tension. Rupture of the interface was clearly visible for tensile load while no obvious indication of mechanical failure was observed under compression. These results suggest that assuming a rigid connection or frictionless sliding contact between the brain tissue and skull bone, the approaches often used in computational biomechanics models of the brain, may not accurately represent the mechanical behaviour of the brain-skull interface.

## 1      Introduction

Computational biomechanics head models implemented using the finite element (FE) method were first reported in the early 1980s [1], and since then, have been a subject of ongoing research effort. Nowadays, comprehensive brain models encompass a wide variety of applications including crash injury prediction and computer-assisted neurosurgery [2-5]. Computational biomechanics brain models include four main elements: (i) the geometry of the brain, typically obtained from electronic brain atlases or magnetic resonance images (MRIs); (ii) boundary and contact conditions; (iii) loading conditions, and (iv) the material laws and material properties of intracranial components. The brain geometry, loading conditions and constitutive properties have been extensively studied. However, there is a lack of experimental evidence to support the currently used models of the brain-skull interface (layers of tissue located between the brain cortical surface and skull inner surface) that determines boundary conditions for the brain [6]. Therefore, parametric studies have been conducted to evaluate the effects of different approaches for modelling the brain-skull interface on the brain responses predicted by computational biomechanics models [6-8].

The brain-skull interface is composed of three main membranous layers (meninges): (i) Dura mater (ii) Arachnoid and (iii) Pia mater [9]. These layers are sandwiched between the rigid cranial inner surface and the brain's outer surface (Fig. 1a). Space between the arachnoid and pia mater (the subarachnoidal space) is filled with cerebrospinal fluid (CSF) and arachnoid trabeculae that are extended from the inner surface of the arachnoid to the pia mater [10, 11]. The exact anatomical structure and mechanical properties of these sublayers are still a matter of debate as scarce quantitative information exists regarding their mechanical properties and interactions [6, 9].

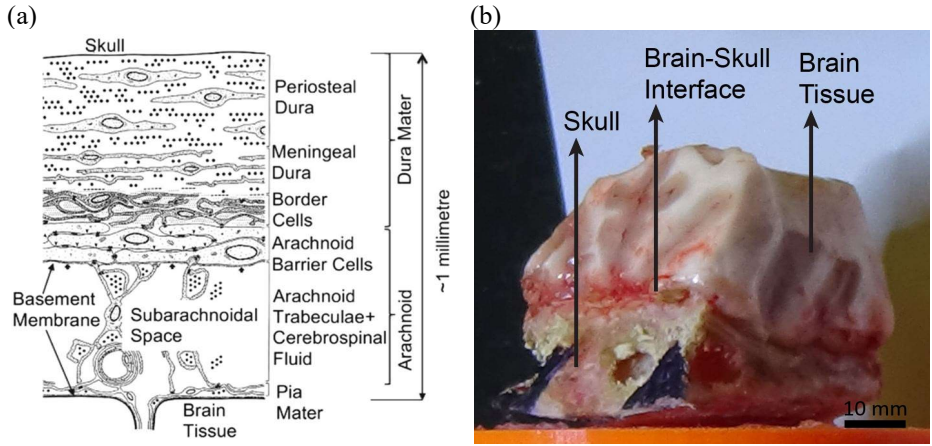

**Fig. 1.** (a) Anatomical structure of the brain-skull interface (meninges). Copied from [6]. (b) Brain-skull complex sample extracted from sheep cadaver head showing the brain tissue, brain-skull interface, and skull bone.

One possible method of investigating the mechanical behaviour of the brain-skull interface involves conducting mechanical tests using samples of tissues that form the brain-skull interface extracted from animal cadaver heads. For instance, Jin et al. [12] investigated the mechanical properties of the pia-arachnoid complex extracted from a cadaver bovine head under in-plane tension, normal traction, and shear loadings. They found that the pia-arachnoid complex exhibits nonlinear viscoelastic material behaviour and is significantly stiffer under in-plane tension than under normal traction and shear. Although studies of meninges and pia-arachnoid tissue samples provide valuable insight into the biomechanical behaviour of the anatomical components of the brain-skull interface, application of their results in comprehensive biomechanical simulations of the brain poses significant challenges due to the risk of inducing damage in thin tissue layers when excising the samples and complexity that would be required to construct the brain-skull interface model that explicitly represents anatomical components of the brain-skull interface and mechanical interactions between these components.

This challenge can be overcome by conducting experiments on samples of the brain-skull complex, which includes skull bone, brain-skull interface, and brain tissues (Fig. 1b). This approach allows the brain-skull interface to remain intact when excising the samples. In 2015, Agrawal et al. [13] conducted compressive tests on the brain-skull complex samples extracted from sheep cadaver heads. They used a finite element model consisting of skull bone and brain tissue with frictionless contact between them to analyse their experiments. Although they used a contact digital scanner to obtain accurate 3D geometry of the skull bone for the model, simplified cuboidal geometry of the brain tissue sample was assumed.

In this study, we demonstrate a new method for investigation of the mechanical properties of the brain-skull interface. The method consists of uniaxial tension and compression experiments on the brain tissue and brain-skull complex samples, with computational biomechanics modelling applied to analyse the experimental results to determine the subject-specific constitutive properties of the brain tissue and mechanical behaviour of the brain-skull interface. Modelling of the experiments was conducted using the finite element method (FEM) and accurate 3D geometry of the samples for the models was obtained from MRIs.

## 2 Methods

### 2.1 Sample Preparation

The experiments, including sample preparation, MRI acquisition, and biomechanical tests, were conducted at the laboratory facilities of the Harry Perkins Institute of Medical Research (Nedlands, WA, Australia). Two sheep cadaver skinned heads were collected from an abattoir (Dardanup Butchering Company) and transported to the laboratory (with The University of Western Australia Biosafety Approval F69199). As the brain-skull complex is a structure (consisting of the brain tissue, brain-skull interface, and skull bone), the subject-specific material properties of brain tissue first needed to be determined. Therefore, biomechanical tests were conducted on two sets of cuboidal samples extracted from each head: (i) brain tissue and (ii) a brain-skull complex consisting of the brain tissue, meninges, and skull bone. Experiments were completed within 24 hours after the animals' death. This is shorter than the delays that could pose a substantial risk of alterations of the brain tissue characteristics [14].

Fig. 2a and Fig. 2b show the sheep cadaver skinned head before and after the sample extraction. To avoid damage to the brain-skull interface, a dental technician drill set (Saeshin Strong 206/H450 by Saeshin, Daegu, Korea) equipped with a Dynex cutting disc (22x0.3 mm) was used to cut the skull. Subsequently, a microtome blade (Erma Patho Cutter HP-R 35/75mm) was used to cut the brain tissue in a dorsal-ventral direction (in sagittal and coronal planes). Fig. 2c and Fig. 2d show the cutting disc and microtome blade, respectively. The cutting lines and the locations of the three extracted parts are shown in Fig. 2a. To minimise the risk of heat- and mechanically induced damage to the brain tissue and brain-skull interface, dissection of skull bone was conducted in two stages:

1) An initial cut that penetrated through most of the bone thickness.
2) A final fine cut at a slow feed speed to prevent the discs from cutting through the brain tissue.

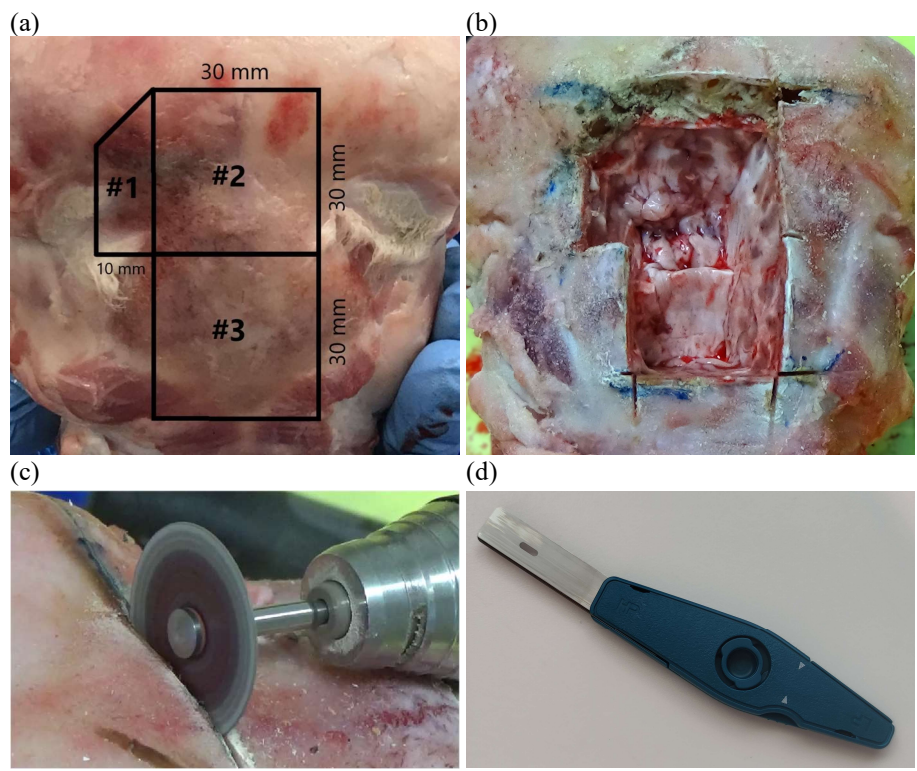

**Fig. 2.** Cadaver sheep head (a) before and (b) after sample extraction. The cutting lines and part numbers are illustrated in (a). (c) Dental technician drill equipped with cutting disc. (d) Microtome blade.

Part 1 was removed first to allow access to the area beneath Part 2 and to cut it horizontally (in the transverse plane). Part 2 was used as the brain-skull complex sample. After the extraction of Part 2, the area beneath Part 3 became accessible. Part 3 was used as the brain tissue sample following the removal of the skull bone. In total, one brain tissue sample (intended dimensions of around 27 mm x 27 mm x 17 mm) and one brain-skull complex sample (intended brain tissue dimensions of around 26 mm x 27 mm x 17 mm) were extracted from each cadaver sheep head. The T2-weighted MRIs of all the samples were then acquired. The samples were not frozen at any time and were kept hydrated inside containers filled with 100 g of 5% saline solution.

## 2.2    Acquisition of Sample Images

Magnetic resonance images (MRIs) of the sheep cadaver heads and tissue samples were acquired using Bruker Biospec 9.4 Tesla animal scanner at the Western Australia node of the National Imaging Facility (NIF) located at the Harry Perkins Institute of Medical Research (Nedlands, WA, Australia). We used a 3D T2-weighted TurboRARE sequence (repetition time = 2200 ms, echo time = 34 ms, RARE factor = 12, partial Fourier acceleration factor = 1.4, matrix size 396x192x48), resulting in a final resolution of 0.25 mm x 0.25 mm x 0.5 mm.

## 2.3    Experiments

The UniVert (CellScale, www.cellscale.com), a specialised portable biomaterial testing system driven by a stepper motor under closed loop control, was used for uniaxial compression and tensile tests. The testing system was equipped with a 10 N LFT-25-1KG load-cell (accuracy of 0.2% of the maximum load, which implies 0.02 N) and displacement transducer. The force and loading head displacements were recorded with a sampling rate of 100 Hz. The experimental protocol was based on Miller [15, 16] and Miller and Chinzei [17, 18]. Dorsal (cortical) surface of the brain tissue samples and outer skull surface for the brain-skull complex samples were glued to the UniVert testing system platen using a fast-curing cyanoacrylate glue (Fix&Go Supaglu by Selleys'). The loading was applied to the ventral surface of the samples. Following Miller [15, 16] and Miller and Chinzei [17, 18], sandpaper (80 grade) was glued to the loading head when subjecting the brain tissue and brain-skull complex samples to compression. This implied no-slip boundary conditions on the sample surface in contact with the loading head in the compressive tests, i.e. movement of the top surface of the samples in such tests was constrained to vertical direction.

In the tensile tests, the ventral surface of the sample was glued to the loading head using the same fast-curing cyanoacrylate glue that was used to attach the sample's dorsal surface to the bottom platen of the UniVert testing system. To ensure adhesion, the loading head was moved 1 mm downwards beyond its initial contact point with the ventral surface of the samples. It was held in this position for 20 s, then returned to the original contact point and held there for additional 60 s to provide stress relaxation before the start of the tensile tests. This procedure follows Miller [16].

In all experiments conducted in this study, the loading speed was 0.3 mm/s which implies a nominal strain rate of approximately 0.02/s.

All tests were conducted at room temperature.

The experiments were recorded using two digital cameras: 1) High resolution DFK 33UX264 scientific camera (resolution of 2048x2048, 15 Frames/s) that provided the frontal view of the test samples; and 2) Sony HDR-CX625 camcorder (resolution of 1920x1080, 25 Frames/s) that provided the general oblique view of the test samples.

The brain tissue and brain-skull complex samples extracted from the first sheep cadaver head were subjected to tension and referred to as H1B-T and H1S-T, respectively. The samples extracted from the second sheep cadaver head were subjected to compression and referred to as H2B-C and H2S-C, respectively. Fig. 3 shows the samples in initial (unloaded) configuration before starting the mechanical tests.

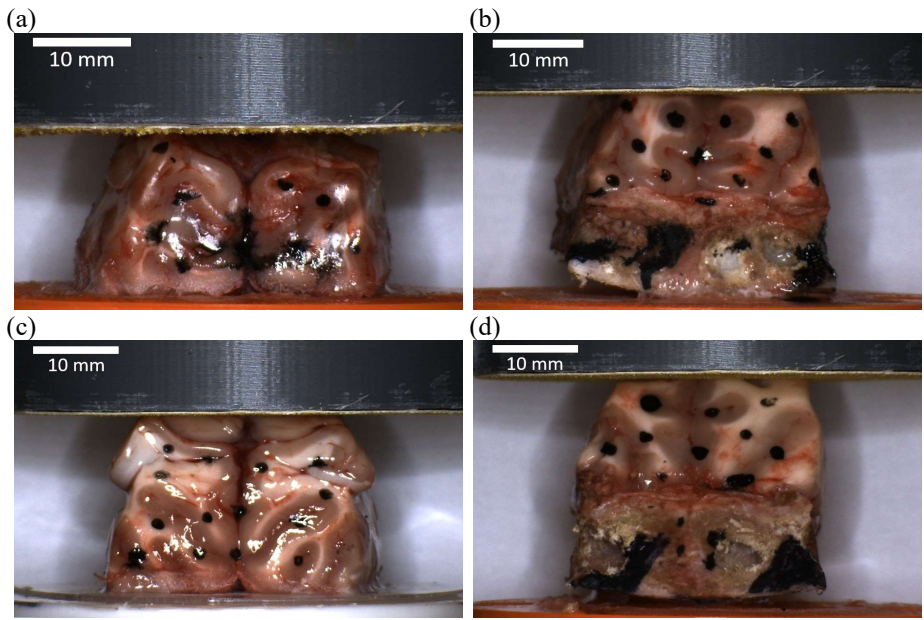

**Fig. 3.** The samples before starting the tests. (a) brain tissue sample used in tensile test (H1B-T), (b) brain-skull complex used in tensile test (H1S-T), (c) brain tissue sample used in compressive test (H2B-C), and (d) brain-skull complex used in compressive test (H2S-C).

### 2.4 Modelling

#### 2.4.1 Finite element mesh generation

3D Slicer software [19] was used to semi-automatically segment the MRIs of the samples to extract their accurate 3D geometry. 3D surfaces constructed from the MRI segmentations using 3D Slicer software were used as a starting point to create finite element meshes of the samples. Meshing was done using Coreform Cubit 2024.3 (Coreform, www.coreform.com/products/coreform-cubit) finite element mesh generator and imported into Abaqus 2023 finite element software (Dassault Systèmes, www.3ds.com/products/simulia/abaqus). This allowed us to create high-quality (mean

Jacobian between 0.62 and 0.75) fully hexahedral (using 8-noded elements) meshes of the samples (Fig. 4 and Table 1).

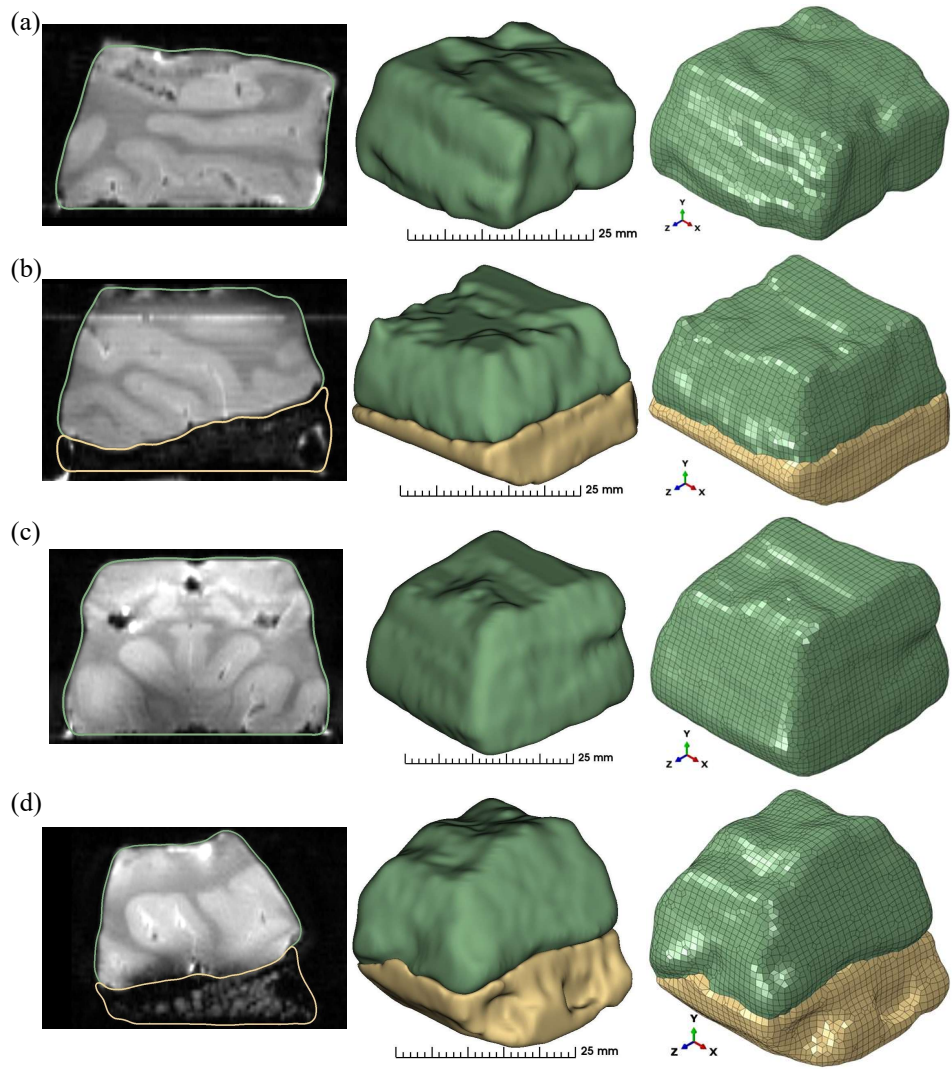

**Fig. 4.** The MRI of the samples with highlighted boundary of material segmentation (left column), 3D representation of the segmentation (middle column), and meshed geometries imported into Abaqus (right column). Brain: green, and skull: yellow. (a) brain tissue sample used in tensile test (H1B-T), (b) brain-skull complex used in tensile test (H1S-T), (c) brain tissue sample used in compressive test (H2B-C), and (d) brain-skull complex used in compressive test (H2S-C).

**Table 1.** The finite element mesh metrics of the brain tissue in samples.

| Sample | Number of nodes | Number of elements | Mean Jacobian |
| --- | --- | --- | --- |
| H1B-T | 16938 | 14969 | 0.6745 |
| H1S-T | 15212 | 13376 | 0.6246 |
| H2B-C | 21903 | 19658 | 0.7504 |
| H2S-C | 16657 | 14699 | 0.6815 |

### 2.4.2 Finite element solver and element formulation

All simulations were conducted using the Abaqus explicit dynamics non-linear finite element solver. We used stiffness-based hourglass control and automated time stepping. To prevent volumetric locking, we used under-integrated hexahedral element formulation (C3D8R element type in Abaqus).

### 2.4.3 Boundary conditions and loading

In all tests of the brain tissue samples (tension: sample H1B-T; compression: sample H2B-C), the bottom (dorsal) surfaces of the samples were glued to the UniVert testing system platen. Hence, in the FE models of these tests, the bottom nodes of the samples were rigidly constrained. In all tests of the brain-skull complex samples (tension: sample H1S-T; compression: sample H2S-C), the outer surface of the skull bones was glued to the UniVert testing system platen. Hence, in the FE models of such tests, the skull bones were rigidly constrained.

In the tensile tests, the top surface of the sample was bonded to the loading head using adhesive. In the compressive tests, the sandpaper glued to the loading head constrained movement of the sample surface in contact with the loading head to vertical direction. Therefore, when modelling the tensile and compressive tests, loading was defined by prescribing the displacement (rate of 0.3 mm/s) in the vertical direction on the model nodes representing the top surfaces of the samples. The displacement was applied using a smooth step procedure in the Abaqus finite element software.

In the FE models of the brain-skull complex samples (tension: sample H1S-T; compression: sample H2S-C), the brain-skull interface was modelled as a rigid tie between the brain tissue and skull. This approach has been used in several previous studies [6, 20].

### 2.4.4 Constitutive models and parameters

The skull was modelled as rigid since it is orders of magnitude stiffer than the brain tissue. Following Miller [15, 16] and Miller and Chinzei [17, 18], we used a first-order Ogden-type hyperelastic model to describe the material behaviour of the brain tissue:

$$W(\lambda_1, \lambda_2, \lambda_3) = \frac{2\mu}{\alpha^2}(\lambda_1^{\alpha} + \lambda_2^{\alpha} + \lambda_3^{\alpha} - 3) + \frac{1}{D}(J_{el} - 1)^2, \qquad (1)$$

where $\lambda_1$, $\lambda_2$ and $\lambda_3$ are the principal stretches, $J_{el}$ is the third strain invariant (also referred to as the volumetric strain), $\mu$ is the shear modulus, $\alpha$ is the dimensionless constant, and $D$ determines the compressibility of the material:

$$D = \frac{2}{K}, \qquad (2)$$

where $K$ is the bulk modulus. The relationship between the bulk modulus, shear modulus, and Poisson's ratio $\nu$ is:

$$\nu = \frac{3\frac{K}{\mu} - 2}{6\frac{K}{\mu} + 2}. \qquad (3)$$

As there is consensus in the literature that the brain and other soft tissues are nearly incompressible [21], we used Poisson's ratio of 0.49 for the brain tissue. When determining the subject-specific brain material constants $\mu$ and $\alpha$ of the brain tissue, the force-displacement curves from the FE models were compared to those obtained in the experiments. We used an iterative process in which the constants were calibrated to match the force-displacement curves observed in the experiments. Following Miller [17], the calibration was done for uniaxial strain between -0.3 (compression) and 0.2 (tension). This resulted in a unique set of material constants that accurately describe the brain tissue material behaviour under both tension and compression.

The subject-specific material constants were then used when simulating the experiments on the brain-skull complex samples extracted from the same sheep cadaver head to investigate the biofidelity of modelling the brain-skull interface as a rigid tie between the brain tissue and skull.

## 3   Results

Fig. 5 shows the force-displacement curves obtained in the experiments and from the FE models of the brain tissue samples subjected to tension (sample H1B-T) and compression (sample H2B-C). The maximum difference between the force obtained from the models and force measured experimentally was 0.04 N (10% of the experimentally measured force) for tension and 0.15 N (11% of the experimentally measured force) for compression. These values indicate a very close agreement between the modelling and experimental results. Table 2 reports the subject-specific Ogden hyperelastic material constants (shear modulus $\mu$ and dimensionless parameter $\alpha$, see Equation 1) of the brain tissue determined in this study. The dimensionless parameter $\alpha$ we obtained here is very close to that reported by Agrawal et al. [13] for sheep brain tissue. The results reported in Fig. 5 confirm the previous findings [17] that the brain tissue stiffness under tension is appreciably lower than under compression.

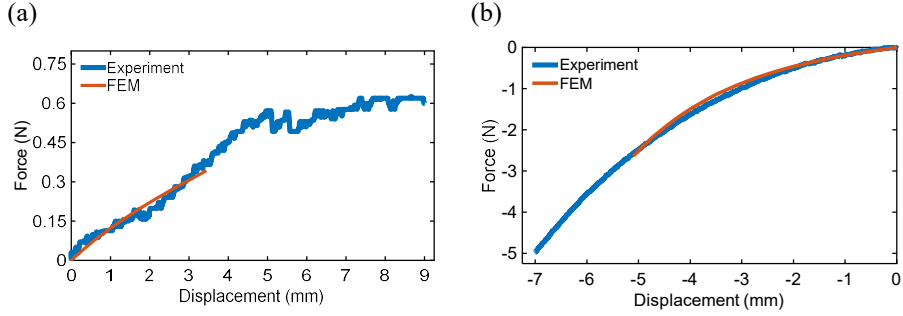

**Fig. 5.** Results for the brain tissue samples. Comparison of the force-displacement curves observed in the experiments and obtained from the FE models when analysing the experimental results to determine the brain tissue material constants for (a) the brain tissue sample subjected to tension (H1B-T); and (b) the brain tissue sample subjected to compression (H2B-C). The term displacement refers to the displacement of the test system loading head. Negative displacement means compression; and positive displacement means tension (sample elongation).

**Table 2.** Ogden hyperelastic material constants of the brain tissue determined in this study through application of the FE models to analyse the experimental results. The constants were determined from the compressive and tensile tests of the brain tissue samples (see Fig. 3a and Fig. 3c) for nominal strain between -0.3 (compression) and 0.2 (tension).

| Material | Loading speed (mm/s) | µ (Pa) | α |
| --- | --- | --- | --- |
| Brain tissue | 0.3 | 1200 | -6.3 |

Fig. 6 shows the force-displacement curves obtained in the experiments and from the FE models of brain-skull complex samples (H1S-T and H2S-C) subjected to compression and tension. Fig. 7 shows these samples in their deformed states under loading. In the tensile test of the brain-skull complex sample (H1S-T), the force continuously increased for the displacement of up to 3.3 mm (see Fig. 6a). It rapidly dropped at 3.3 mm and then started to increase again. We interpret this force drop as an indication of the brain-skull interface mechanical failure and the start of detachment of the brain-skull interface from the skull bone. This detachment initially occurred only locally (see Fig. 7b). Then, it rapidly expanded until the entire interface detached from the skull, with the connection between the brain and skull maintained solely by the bridging veins (see Fig. 7c and Fig. 7d). We attribute the continuous increase in the force magnitude following the sudden drop at 3.3 mm to elongation of the bridging veins and dura mater connected to falx. Elongated bridging veins are clearly visible in Fig. 7d.

As inspection of the skull bone forming the brain-skull complex samples after the experiments indicated that the dura remained intact and firmly attached to the skull (see Fig. 8a), we suggest that the failure of the brain-skull interface may have occurred in the layers spanning arachnoid, arachnoid barrier cells, and border cells (see Fig. 8b).

Modelling of the tensile test of the brain-skull complex (H1S-T sample) using the assumption that the brain is rigidly attached to the skull resulted in a force-displacement relationship that accurately represents the experimental results for the sample elongation of up to 2 mm. This suggests that the tensile stiffness of the brain-skull interface, including the bridging veins, dura, and falx, may be appreciably higher than that of the brain tissue. This suggestion is consistent with the previous findings [22-24].

The key shortcoming of modelling the brain-skull interface as a rigid connection (tie) between the brain and skull is that it does not account for the interface failure observed in our experiments.

The force-displacement relationship of the brain-skull sample complex under tension was nearly linear for the displacement between 2 mm and 3 mm (Fig. 6a). One possible explanation for this behaviour may be the contribution of forces induced in the bridging veins and falx. The effects of the bridging veins and falx may also be a possible explanation for the relatively large difference (0.23 N, 30% of the experimentally measured force) between the experimental and modelling results for the brain-skull sample under tension (sample H1S-T).

For compression, the force-displacement curve of the brain-skull complex sample (H2S-C) (Fig. 6b) exhibited a similar trend to that of the brain tissue sample (H2B-C) (Fig. 5b). For the sample top surface displacement of up to around 4 mm (nominal compressive strain of around -0.23), the modelling results agreed well with the experiments. For the displacement exceeding 4 mm compression, the model tended to overestimate the forces measured in the experiments. For 5.1 mm compression, the difference the experimental and modelling results was 0.59 N (20% of the experimentally measured force). It may be tempting to explain this discrepancy by possible damage of the brain-skull interface and/or brain tissue for the compressive displacement of the sample top surface exceeding 4 mm. However, this interpretation of the results reported in Fig. 6b should be treated very carefully as no obvious indication of such damage under compressive loading was visually observed here (see Fig. 7e and Fig. 7f).

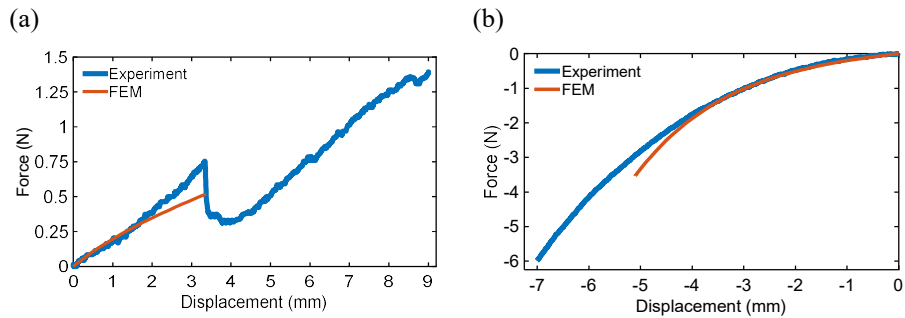

**Fig. 6.** Force-displacement curves for the samples of the brain–skull complex observed in the experiments and obtained from the FE models of the experiments. (a) Brain-skull complex sample subjected to tension (H1S-T); and (b) Brain-skull complex sample subjected to compression (H2S-C). In FE models, the brain tissue was rigidly attached to the skull bone. Positive displacement means tension (sample elongation) and negative displacement means compression.

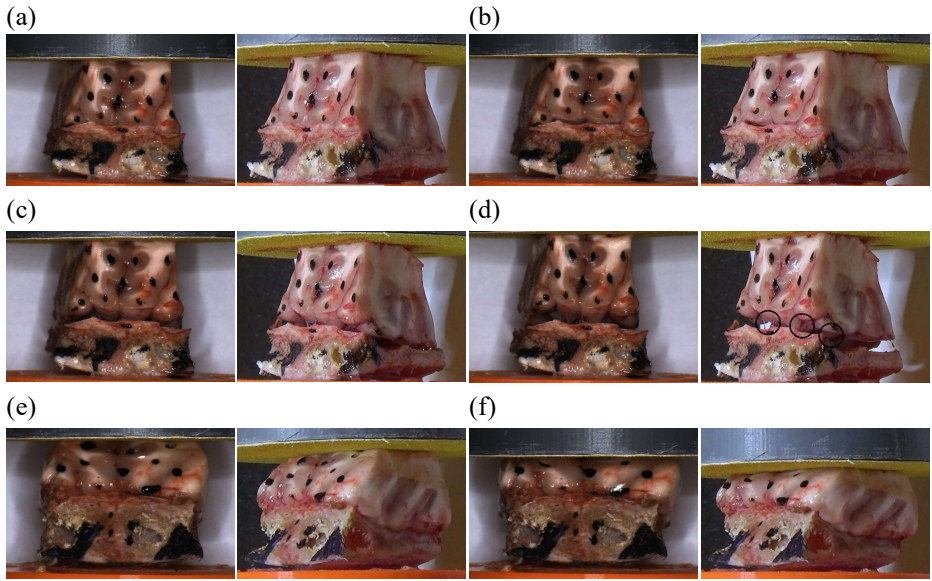

**Fig. 7.** Front and oblique views of the brain-skull complex samples in the experiments conducted in this study. (a) 3.3 mm elongation; (b) 3.6 mm elongation; (c) 6 mm elongation; (b) 9 mm elongation; (e) 3 mm compression; and (f) 6 mm compression. The load (displacement) was applied to the top (ventral) surfaces of the samples. Note detachment of the brain tissue from the skull for 6 mm elongation (c) and 9 mm elongation (d). The elongated bridging veins are indicated in (d).

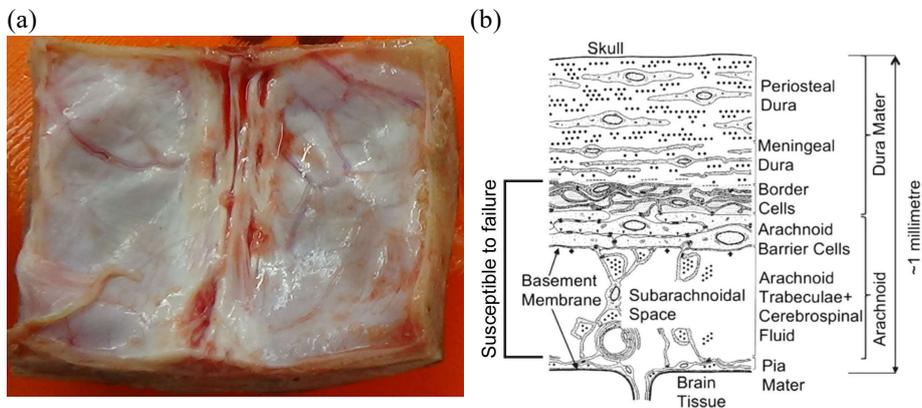

**Fig. 8.** (a) The dura remained intact and firmly attached to the skull after failure of the brain-skull interface under tension. (b) Anatomical structure of the brain-skull interface (meninges) with an indication of tissue layers that may be susceptible layers to failure under tensile load. Modified from [6].

## 4   Discussion

In this study, we investigated the mechanical properties of the brain-skull interface under compression and tension. Brain tissue and brain-skull complex samples (consisting of brain tissue, brain-skull interface, and skull bone) were extracted from sheep cadaver heads and subjected to uniaxial compression and tension. Subject-specific constitutive parameters of the brain tissue (Ogden constitutive model was used) were obtained through comprehensive finite element (FE) modelling of the experiments using the sample geometry obtained from magnetic resonance images (MRIs). The results are consistent with the literature and indicate that the brain tissue is appreciably stiffer under compression than tension.

The subject-specific material properties of the brain tissue were used in comprehensive modelling of the experiments on the brain-skull complex samples extracted from the same head to analyse the biomechanical behaviour of the brain-skull interface. We found that rigidly connecting the brain to the skull, one of the approaches for modelling boundary conditions for the brain used in the literature, may not be adequate as it does not capture the brain-skull interface failure under tension observed in the experiments conducted in the study. Our experiments point also to the limitations of modelling the interactions between the brain and skull using a frictionless sliding contact, an approach used in our previous studies on neurosurgery simulation [3, 25-28], as it allows free separation of the brain cortical surface from the skull.

In this study, we focused on analysing the overall behaviour of the brain-skull interface. While we acknowledge the importance of cerebrospinal fluid (CSF) in influencing this behaviour, quantifying the effects of CSF would require further studies.

It can be concluded that this study confirms the feasibility of the proposed method for determining the biomechanical behaviour of the brain-skull interface using uniaxial compression and tension experiments combined with comprehensive modelling of the experiments and provides preliminary insight into such behaviour. However, it also highlights the need for conducting experiments on a larger number of brain tissue and brain-skull complex samples and indicates that an accurate representation of the biomechanical behaviour of the brain-skull interface would require models more advanced than those currently used in the literature. In particular, accounting for the mechanical failure of the brain-skull interface under tension may be one of the requirements such models need to satisfy.

**Acknowledgement.** This research was supported by the Australian Government through the Australian Research Council's ARC Discovery Projects funding scheme (project DP230100949). Author S. A. acknowledges support of the University of Western Australia Scholarship for International Research Fees and University of Western Australia (UWA) – Intelligent Systems for Medicine Laboratory (ISML) Higher Degree by Research Scholarship in Computational Biomechanics (partly funded from Australian Research Council's ARC Discovery Project DP230100949). The authors acknowledge the facilities and scientific and technical assistance of the National Imaging Facility, a National Collaborative Research Infrastructure Strategy (NCRIS) capability, at the Centre for Microscopy, Characterisation and Analysis, The University of Western Australia. The authors thank Mr Bernard Panizza of Dardanup Butchering Company (Bunbury, Western Australia, Australia) for help in obtaining sheep cadaver heads for the experiments conducted in this study.

## References

1. Shugar, T.A. and M.G. Katona, *Development of finite element head injury model.* Journal of the Engineering Mechanics Division, 1975. **101**(3): p. 223-239.
2. Matsuda, T., et al. *Development of a human body model (THUMS Version 7) to simulate kinematics and injuries of reclined occupants in frontal collisions.* in *27th International Technical Conference on the Enhanced Safety of Vehicles (ESV) National Highway Traffic Safety Administration.* 2023.
3. Miller, K., et al., *Biomechanical modeling and computer simulation of the brain during neurosurgery.* International Journal for Numerical Methods in Biomedical Engineering, 2019. **35**(10): p. e3250.
4. Carmo, G.P., et al., *Development, validation and a case study: The female finite element head model (FeFEHM).* Computer Methods and Programs in Biomedicine, 2023. **231**: p. 107430.
5. Lu, Y.C., et al., *A 3D computational head model under dynamic head rotation and head extension validated using live human brain data, including the falx and the tentorium.* Annals of Biomedical Engineering, 2019. **47**: p. 1923-1940.
6. Wang, F., et al., *Prediction of brain deformations and risk of traumatic brain injury due to closed-head impact: quantitative analysis of the effects of boundary conditions and brain tissue constitutive model.* Biomechanics and Modeling in Mechanobiology, 2018. **17**: p. 1165-1185.
7. Yang, S., et al., *Assessment of brain injury characterization and influence of modeling approaches.* Scientific Reports, 2022. **12**(1): p. 13597.
8. Zhou, Z., X. Li, and S. Kleiven, *Evaluation of brain-skull interface modelling approaches on the prediction of acute subdural hematoma in the elderly.* Journal of Biomechanics, 2020. **105**: p. 109787.
9. Haines, D.E., H.L. Harkey, and O. Al-Mefty, *The "subdural" space: a new look at an outdated concept.* Neurosurgery, 1993. **32**(1): p. 111-120.
10. Walsh, D.R., et al., *Mechanical properties of the cranial meninges: a systematic review.* Journal of Neurotrauma, 2021. **38**(13): p. 1748-1761.
11. Mack, J., W. Squier, and J.T. Eastman, *Anatomy and development of the meninges: implications for subdural collections and CSF circulation.* Pediatric Radiology, 2009. **39**: p. 200-210.
12. Jin, X., et al., *Constitutive modeling of pia–arachnoid complex.* Annals of Biomedical Engineering, 2014. **42**: p. 812-821.
13. Agrawal, S., et al. *Mechanical properties of brain–skull interface in compression.* in *Computational Biomechanics for Medicine: New Approaches and New Applications.* 2015. Springer.

14. Rashid, B., M. Destrade, and M.D. Gilchrist, *Influence of preservation temperature on the measured mechanical properties of brain tissue.* Journal of Biomechanics, 2013. **46**(7): p. 1276-1281.
15. Miller, K., *Method of testing very soft biological tissues in compression.* Journal of Biomechanics, 2005. **38**(1): p. 153-158.
16. Miller, K., *How to test very soft biological tissues in extension?* Journal of Biomechanics, 2001. **34**(5): p. 651-657.
17. Miller, K. and K. Chinzei, *Mechanical properties of brain tissue in tension.* Journal of Biomechanics, 2002. **35**(4): p. 483-490.
18. Miller, K. and K. Chinzei, *Constitutive modelling of brain tissue: experiment and theory.* Journal of Biomechanics, 1997. **30**(11-12): p. 1115-1121.
19. Fedorov, A., et al., *3D Slicer as an image computing platform for the Quantitative Imaging Network.* Magnetic Resonance Imaging, 2012. **30**(9): p. 1323-1341.
20. Claessens, M., F. Sauren, and J. Wismans, *Modeling of the human head under impact conditions: a parametric study.* SAE Transactions, 1997: p. 3829-3848.
21. Miller, K., *Biomechanics of the Brain*. 2011: Springer.
22. Mazumder, M., et al., *Mechanical properties of the brain–skull interface.* Acta of Bioengineering and Biomechanics, 2013. **15**(2): p. 3-11.
23. Famaey, N., et al., *Structural and mechanical characterisation of bridging veins: A review.* Journal of the Mechanical Behavior of Biomedical Materials, 2015. **41**: p. 222-240.
24. Walsh, D.R., et al., *Mechanical characterisation of the human dura mater, falx cerebri and superior sagittal sinus.* Acta Biomaterialia, 2021. **134**: p. 388-400.
25. Wittek, A., et al., *Patient-specific non-linear finite element modelling for predicting soft organ deformation in real-time; Application to non-rigid neuroimage registration.* Progress in Biophysics and Molecular Biology, 2010. **103**(2-3): p. 292-303.
26. Miller, K., et al., *Beyond finite elements: a comprehensive, patient-specific neurosurgical simulation utilizing a meshless method.* Journal of Biomechanics, 2012. **45**(15): p. 2698-2701.
27. Joldes, G.R., A. Wittek, and K. Miller, *Suite of finite element algorithms for accurate computation of soft tissue deformation for surgical simulation.* Medical Image Analysis, 2009. **13**(6): p. 912-919.
28. Joldes, G.R., et al. *Realistic and efficient brain-skull interaction model for brain shift computation*. in *Computational Biomechanics for Medicine III Workshop, MICCAI*. 2008.